\documentclass[12pt]{article}

\begin{document}

\begin{center}
\LARGE{Explanation of the mass of the muon}

\bigskip

\Large{E. L. Koschmieder}
\medskip

\small{Center for Statistical Mechanics, The University of Texas at 
Austin\\Austin, TX 78712, USA
\\email:koschmieder@mail.utexas.edu}
\smallskip
\end{center}

\bigskip
\noindent
\small{The difference of the rest masses m($\pi^\pm) - \mathrm{m}(\mu^\pm$) is 
nearly equal to 1/4 of the rest mass of the $\pi^\pm$ mesons and is equal
 to the sum of the rest masses of the 0.7$\cdot 10^9$ muon neutrinos 
(respectively anti-muon neutrinos) which are in the cubic lattice of the 
$\pi^\pm$ mesons according to the standing wave model. In the decay of a 
$\pi^+$ or $\pi^-$ meson all muon neutrinos, respectively anti-muon 
neutrinos, of the cubic lattice of the $\pi^\pm$ mesons are emitted. The sum
 of the oscillation energies of all 
neutrinos in the $\pi^\pm$ mesons is the same as the sum of the 
oscillation energies of the remaining neutrinos in the $\mu^\pm$ mesons. 
Consequently the mass of the $\mu^\pm$ mesons is equal to m($\pi^\pm) - 
0.7\cdot 10^9 \,\mathrm{m}(\nu_\mu)\,\approx \, 0.75\hspace{.1cm}\mathrm{m}
(\pi^\pm)$, within 1\% in agreement with the measured ratio 
$\mathrm{m}(\mu^\pm)\, / \,\mathrm{m}(\pi^\pm)$ = 0.757028.}

\normalsize

\section{Introduction}
In a previous paper [1] we have found that the ratios of the masses of the 
so-called stable mesons and baryons of the $\gamma$-branch of the spectrum 
of the stable elementary particles are integer multiples of the mass of 
the $\pi^0$  meson within, on the average, 0.73\%. We have explained this 
empirical fact with the standing wave model [2]. The ratios of the masses 
of the neutrino branch of the stable particle spectrum are integer 
multiples of the mass of the $\pi^\pm$ mesons times a common factor 
$0.86\,\pm\,0.02$ [1]. We have explained the ratios of the masses of the 
particles of the $\nu$-branch with the standing wave model in [3]. 
Surprisingly we can also explain the ratio of the mass of the $\mu^\pm$ 
mesons to the mass of the $\pi^\pm$ mesons with the standing wave model.

\section{The mass of the muons}
The mass of the $\mu^\pm$ mesons is m($\mu^\pm$) = 105.658389 $\pm\,  
3.4\cdot10^{-5}$ MeV, according to the Particle Physics Summary [4]. The 
muons are stable, their lifetime is $\tau_\mu = 
2.19703\cdot10^{-6}\pm4\cdot10^{-11}$ sec, about a hundred times as long 
as the lifetime of the $\pi^\pm$ mesons, that means longer than the 
lifetime of any other elementary particle, but for the electrons, protons 
and neutrons. The muons are part of the lepton family which is 
distinguished from the mesons and baryons not so much by their mass as the 
name lepton implies, actually the mass of the $\tau$ meson is about twice 
the mass of the proton, but rather by the absence of strong interaction 
with the mesons and baryons. The masses of the leptons are not 
explained by the standard model of the particles.

   Comparing the mass of the $\mu^\pm$ mesons to the mass of the $\pi^\pm$ 
mesons m($\pi^\pm$) = 139.56995 MeV we find that m($\mu^\pm$)/m($\pi^\pm$) 
= 0.757028 $\approx3/4$ or that m($\pi^\pm) - \mathrm{m}(\mu^\pm$) = 33.91156 
MeV = 0.24297$\cdot\mathrm{m}(\pi^\pm)$ or $\approx 1/4 \,\cdot$ m$(\pi^\pm$). The mass of the electron is approximately 
1/206 of the mass of the muon, the contribution of m(e) to m($\mu$) will 
therefore be neglected in the following. We assume, as we have done in [2] 
and [3] and as appears to be natural, that the particles, including the 
muons, consist of the particles into which they decay. The $\mu^+$ meson 
decays via $\mu^+\rightarrow e^+ +\bar{\nu}_\mu+\nu_e$. The muons are 
apparently composed primarily of the neutrinos which are already present 
in the cubic neutrino lattice of the $\pi^\pm$ mesons according to the 
standing wave model [3]. The $\pi^+$ meson decays via 
$\pi^+\rightarrow\mu^++\nu_\mu$ or the conjugate particles in the decay of 
the $\pi^-$ meson, with 99.988\% of the $\pi^\pm$ decays in this form. The 
energy m($\pi^\pm) - \mathrm{m}(\mu^\pm$) $\approx 1/4\,\cdot$ m$(\pi^\pm)$ is lost when a 
$\mu^+$ meson and a muon neutrino $\nu_\mu$, respectively a $\mu^-$ and an anti-muon 
neutrino $\bar{\nu}_\mu$, are emitted by the $\pi^\pm$ mesons. The rest of 
the energy in the rest mass of the $\pi^\pm$ mesons passes to the rest 
mass of the $\mu^\pm$ mesons. In the standing wave model of the particles 
of the neutrino branch [3] the $\pi^\pm$ mesons are composed of a cubic 
lattice consisting of N/2 = 0.71$\cdot10^9$ muon neutrinos $\nu_\mu$ 
and the same number of anti-muon neutrinos $\bar{\nu}_\mu$, (m($\nu_\mu$) 
= m($\bar{\nu}_\mu$)), as well as of N/2 electron neutrinos $\nu_e$ and 
the same number of anti-electron neutrinos $\bar{\nu}_e$, (m($\nu_e$) = 
m($\bar{\nu}_e$)), plus the oscillation energy of these neutrinos. We 
found in [3] that the mass of a single muon neutrino should be 50 
milli-eV/$c^2$, and the mass of a single electron neutrino should be 5 
meV/$c^2$. 
Using these values of N and of the neutrino masses we find two intriguing 
``coincidences".
\bigskip

(a) The difference of the rest masses of the $\pi^\pm$ and $\mu^\pm$ mesons 
is nearly equal to the sum of the rest masses 
of all muon neutrinos respectively anti-muon neutrinos in the 
$\pi^\pm$ mesons.

\smallskip
    m($\pi^\pm) - \mathrm{m}(\mu^\pm$) = 33.912 MeV \quad versus \quad                     
N/2$\,\cdot\, \mathrm{m}(\nu_\mu$) = 35.675 MeV.
\bigskip

(b) The energy in the oscillations of all neutrinos in the              
$\pi^\pm$ mesons is nearly the same as the energy in the            
oscillations of all $\bar{\nu}_\mu$, respectively $\nu_\mu$, and 
    $\nu_e$ and $\bar{\nu}_e$ neutrinos in the $\mu^\pm$ mesons.

\smallskip
    $E_{osc}(\pi^\pm$) = m($\pi^\pm) - N[\mathrm{m}(\nu_\mu) +                  
\mathrm{m}(\nu_e$)] = 61.09 MeV \quad versus \hspace{1.5cm}

    $E_{osc}(\mu^\pm) = \mathrm{m}(\mu^\pm) - N/2\,\cdot\mathrm{m}(\bar{\nu}_\mu) - 
    N\mathrm{m}(\nu_e$) = 62.85 MeV.
\bigskip

Both statements are, of course, valid only within the accuracy with which 
the number 2N of all neutrinos in the $\pi^\pm$ lattice is known, as well 
as within the accuracy with which the masses m($\nu_\mu$) and m($\nu_e$) 
have been determined in [3]. It cannot be expected that this accuracy is 
better than a few percent, considering in particular the uncertainty of 
the lattice constant.

   In the two-body decay of the $\pi^\pm$ mesons each decay product has 
the same and specific momentum p which is p = 30 MeV/c in the rest frame, 
according to the Particle Physics Summary [4]. The momentum of the decay 
particles translates into a kinetic energy which for the $\mu$ mesons is 
$E_k(\mu^\pm)$ = 4.1 MeV, and for the single emitted muon neutrino with a 
practically vanishing rest mass $E_k(\nu_\mu)$ = 30 MeV. Hence the kinetic 
energies of the two decay products of the $\pi^\pm$ meson is 34.1 MeV, 
which is, as must be, practically equal to m($\pi^\pm) - \mathrm{m}(\mu^\pm)$ = 
33.9 MeV. In terms of our model of the neutrino lattice in the $\pi^\pm$ 
mesons and the statement (a) above this means that the rest masses of all 
muon neutrinos in the $\pi^\pm$ mesons are converted into kinetic energy. 
All but one of the $0.7\cdot10^9$ muon neutrinos disappear in the decay of the 
cubic neutrino lattice of the $\pi^\pm$ mesons. We learn from the high 
energy collision $e^+ + e^-\rightarrow\mu^+ + \mu^-$ that $\mu$ mesons, 
and the neutrinos which are part of the $\mu$ meson masses, can be created 
directly out of the kinetic energy of the electrons and positrons in the 
$e^+ + e^-$ collision. What we observe in the $\pi^\pm$ decay is the 
reverse process, the conversion of the energy in neutrino rest masses into 
kinetic energy.

   We note that the difference of the rest masses 
m($\pi^\pm) - \mathrm{m}(\mu^\pm)$ provides an independent check of the value of 
the rest mass of the muon neutrino. Using N/2 = 0.7135$\cdot10^9$ \,and 
m($\pi^\pm) - \mathrm{m}(\mu^\pm)$ = 33.912 MeV\, it follows that the rest mass of 
the muon neutrino should be m($\nu_\mu)$ = 47.53 meV/$c^2$, whereas we 
found m($\nu_\mu)$ = 50 meV/$c^2$ in [3].

   Since the decay of the $\pi^\pm$ mesons seems to mean the removal of all 
$\nu_\mu$, respectively, all $\bar{\nu}_\mu$ neutrinos from the neutrino 
lattice of the $\pi^\pm$ mesons, the muons should contain the 
remaining neutrinos of the original cubic lattice, that means N/2 
anti-muon neutrinos $\bar{\nu}_\mu$, respectively, N/2 muon neutrinos 
$\nu_\mu$, plus N/2 electron neutrinos $\nu_e$ as well as N/2 
anti-electron neutrinos $\bar{\nu}_e$. It can be shown immediately that the 
concept that the muons consist of three types of neutrinos and 
their oscillation energies leads to the correct mass of the muons. 
The energy E in the rest mass of the $\mu$ mesons must be equal to the 
oscillation or average kinetic energy of all neutrinos in the particle plus the 
energy in the rest masses of the neutrinos. From E = $\mathrm{E}_k$ + 
$\Sigma \mathrm{m}(\nu)c^2$ follows the elementary formula Eq.(18) in [3], which 
applied to the case of the muons is

\begin{equation}\frac{E(\mu^\pm)}{E(\pi^\pm)} = \frac{E_k(\mu^\pm)} 
{E_k(\pi^\pm)}\cdot\frac{1+\Sigma \mathrm{m}^\prime c^2/E_k(\mu^\pm)}{1+ 
\Sigma \mathrm{m}c^2/E_k(\pi^\pm)}\,. 
\end{equation}

\smallskip
\noindent
With the empirical $E_k(\mu^\pm)=\mathrm{m}(\mu^\pm)_0 c^2-\Sigma \mathrm{m}^\prime c^2$ = 62.9 MeV 
and the empirical $E_k(\pi^\pm)=\mathrm{m}(\pi^\pm)_0 c^2-\Sigma \mathrm{m}c^2$ = 61.1 MeV, where 
$\Sigma \mathrm{m}^\prime c^2=\mathrm{N}/2\,\cdot\,50$ 
meV\,+\,$\mathrm{N}\,\cdot 5$ meV (the sum of the energies of 
the rest masses of the anti-muon neutrinos, electron neutrinos and 
anti-electron neutrinos in the $\mu^+$ meson), and 
$\Sigma \mathrm{m}c^2=\mathrm{N}\,\cdot\,50$ meV $\,+\,\mathrm{N}\,\cdot\, 
5$ meV (the sum of the energies of the 
rest masses of the four neutrino types in the $\pi^\pm$ mesons), it 
follows that

\begin{equation}E(\mu^\pm)/E(\pi^\pm)=0.75763\,,\end{equation}
\noindent
whereas the measured ratio is $E(\mu^\pm)/E(\pi^\pm)$ = 0.757028. That means 
that the concept of the muons consisting of three neutrino types 
plus their oscillation or kinetic energy leads to the correct ratio of the 
mass of the $\mu^\pm$ mesons to the mass of the $\pi^\pm$ mesons.

\section{The oscillation energies of the neutrinos}

The neutrinos in the body of a muon must oscillate because the 
collision $e^+ + e^- \rightarrow \mu^+ + \mu^-$ tells that a continuum of 
frequencies must be present in the muons, if Fourier analysis 
holds. The continuum of frequencies from Fourier analysis can be absorbed 
by the continuous spectrum of the oscillations in a neutrino lattice. The 
energy of the oscillations of the neutrinos in the muon lattice is 
the sum of the energies of a plane oscillation in an 
isotropic, diatomic lattice consisting of N/2 $\bar{\nu}_\mu$ 
neutrinos and N/2 $\nu_e$ neutrinos, part of the remains of the diatomic 
neutrino lattice of the $\pi^\pm$ mesons, plus the energy of the diatomic 
oscillations of N/2 $\bar{\nu}_\mu$ and N/2 
$\bar{\nu}_e$ neutrinos which neutrinos were likewise in the lattice of the 
$\pi^\pm$ mesons. The latter oscillations are likely to be perpendicular to the 
first mentioned diatomic 
oscillations in the muon, because the $\bar{\nu}_e$,\,$\nu_e$ 
neutrino pairs are oriented perpendicular to the $\bar{\nu}_\mu$, 
$\bar{\nu}_e$ neutrino pairs 
in the original cubic lattice of the $\pi^\pm$ mesons, see Fig. 1 of [3].

 The energy of the oscillations of the lattice is given by

\begin{equation}E_{osc}=\frac{Nh\nu_0}{(2\pi)^2}\int\int_{-\pi}^{\pi}f(\phi_1,\phi_2)\, d\phi_1\,d\phi_2 \,,\end{equation}
\noindent
as in Eq.(17) of [3] or in the original paper of Born and v.Karman [5], 
Eq.(50) therein. The oscillation energy of the diatomic lattice containing the 
$\bar{\nu}_\mu$ and $\nu_e$ neutrinos of the $\mu^\pm$ mesons is 1/2 of 
the oscillation energy of the diatomic lattice oscillations in the 
$\pi^\pm$ mesons because the number of the pairs $\bar{\nu}_\mu$ and 
$\nu_e$ in the $\mu^\pm$ mesons is 1/2 of the number of the corresponding 
pairs $\bar{\nu}_\mu$ and $\nu_e$ and $\nu_\mu$ and $\bar{\nu}_e$ in the 
$\pi^\pm$ mesons. The functions f($\phi_1,\phi_2)$ in Eq.(3) describing the 
frequency spectrum of the oscillations are the same for the diatomic 
neutrino pairs in the $\pi^\pm$ mesons and the diatomic neutrino pairs in 
the $\mu^\pm$ mesons. The functions are given by Eq.(6) of [3]. Since both 
functions are the same the ratio of the energy in the diatomic 
$\bar{\nu}_\mu$,\,$\nu_e$ lattice 
oscillation of the $\mu^\pm$ mesons to the energy of the diatomic lattice 
oscillations in the $\pi^\pm$ mesons is = 1/2 according to Eq.(3). But 
since the same applies for the diatomic oscillations of the 
$\bar{\nu}_\mu$,\,$\bar{\nu}_e$ pairs in the $\mu^\pm$ mesons, the sum of the energies of both 
oscillations is equal to the oscillation energy of the $\pi^\pm$ mesons, 
as (b) says.

   Since the frequencies of the diatomic oscillations are a quadratic 
function of $\nu$ there exists for each positive frequency also a 
negative frequency of the same absolute value. That means, according to 
(3), that the sum of the energies of the oscillations with negative 
frequencies is negative, but has the same absolute value as the sum of the 
energies of the oscillations with positive frequencies. That means that 
each muon has an antiparticle. In the antiparticle of a particular muon 
the oscillation energy of the neutrinos is replaced by the oscillation 
energy of the frequencies with the opposite sign, and the masses of the 
neutrinos are replaced by the masses of their antineutrinos, which have 
the same absolute value of the masses they replaced. The energy in the mass 
of a muon is consequently equal to the energy in the mass of its 
antiparticle, as it is with the masses of the particles of the $\gamma$-branch and of the 
$\nu$-branch, and with the masses of the $\mu^+$ and $\mu^-$ mesons. 

   Finally we ask why do the muons not interact strongly with the 
mesons and baryons? In [6] we have shown that a strong force emanates from 
the sides of a cubic lattice. This force is caused by the unsaturated weak forces of 
about $10^6$ neutrinos at a side of the surface of the neutrino lattice of the 
mesons and baryons. The existence of such a force follows from the study of Born and Stern [7] 
which dealt with the forces between two parts of a cubic lattice cleaved 
in vacuum. If the muons have a lattice consisting of 
$\bar{\nu}_\mu$, respectively $\nu_\mu$, and $\nu_e$ and $\bar{\nu}_e$ 
neutrinos the lattice surface is not the same as the surface of the cubic
$\nu_\mu$, $\bar{\nu}_\mu$, $\nu_e$, $\bar{\nu}_e$ 
lattice of the mesons and baryons, as described by the standing wave model 
[2,3]. Therefore it does not seem likely that the muons interact in 
the same way with the mesons and baryons as the mesons and baryons 
interact with each other. To put this in another way, a triclinic lattice 
does not bond with a cubic lattice.

   \section{Conclusions}

It has been shown that the mass of the muons can be explained as 
the sum of the rest masses of $0.71\cdot10^9$  muon neutrinos, respectively 
anti-muon neutrinos, and of the same number of electron neutrinos and 
anti-electron neutrinos, plus their oscillation energies. The three 
neutrino types in a muon are the remains of the cubic neutrino 
lattice of the $\pi^\pm$ mesons from which the muons are formed in 
the $\pi^\pm$ decay. The mass of the muons differs from 
the mass of the $\pi^\pm$ mesons by the sum of the rest masses of all muon 
neutrinos, respectively of all anti-muon neutrinos which are in the 
$\pi^\pm$ mesons. All muon neutrinos of one type are emitted when the 
$\pi^\pm$ mesons decay. Since the sum of the rest masses of all muon 
neutrinos of one type in the $\pi^\pm$ mesons is approximately 1/4 of the rest 
mass of $\pi^\pm$ mesons, and since the oscillation energies in $\pi^\pm$ and 
$\mu^\pm$ are the same, the mass remaining in the $\mu^\pm$ mesons after 
the $\pi^\pm$ decay is $\approx$ 3/4 $\cdot$ m$(\pi^\pm)$, within 1\% in agreement with the 
measured ratio m$(\mu^\pm)$/m$(\pi^\pm)=0.757028$. We have also found that 
the muons do not interact with the mesons and baryons in the same 
strong way as the mesons and baryons interact with each other.

The explanation of the mass of the muons and of their weak interaction is a 
straightforward consequence of our standing wave model of the stable mesons 
and baryons.

\bigskip
\bigskip
\bigskip

\noindent
\textbf{REFERENCES}
\bigskip

\noindent
[1] E.L. Koschmieder, Bull.Acad.Roy.Belgique {\bfseries X}(1999)281,\\
\indent 
 hep-ph/0002179.
\smallskip

\noindent
[2] E.L. Koschmieder and T.H. Koschmieder,
 
Bull.Acad.Roy.Belgique {\bfseries X}(1999)289,\\
\indent
hep-lat/0002016.
\smallskip

\noindent 
[3] E.L. Koschmieder and T.H. Koschmieder, hep-lat/0104016.
\smallskip

\noindent
[4] R.M. Barnett et al., Rev.Mod.Phys. {\bfseries 68}(1996)611.
\smallskip

\noindent
[5] M. Born and Th. von Karman, Phys.Z. {\bfseries 13}(1912)297.
\smallskip

\noindent
[6] E.L. Koschmieder, Nuovo Cim. {\bfseries 101}(1989)1017.
\smallskip

\noindent
[7] M. Born and O. Stern, Sitzber.Preuss.Akad.Wiss. {\bfseries 33}(1919)901.
\smallskip

\end{document}